\documentclass{article}
\usepackage{cite}
\usepackage{hyperref}
\usepackage{spconf,amsmath,graphicx}
\usepackage{url}
\usepackage{multirow} 
\usepackage{authblk}

\begin{document}
\topmargin=0mm
%
%

\name{Chenyang Guo\textsuperscript{\rm 1}, Liping Chen\textsuperscript{\rm 1}, Zhuhai Li\textsuperscript{\rm 1}, Kong Aik Lee\textsuperscript{\rm 2}, Zhen-Hua Ling\textsuperscript{\rm 1}, Wu Guo\textsuperscript{\rm 1}}

\address{
  \textsuperscript{1}University of Science and Technology of China, Hefei, China\\
  \textsuperscript{2}The Hong Kong Polytechnic University, Hong Kong, China
}

\title{On the Generation and Removal of Speaker Adversarial Perturbation for Voice-Privacy Protection}
\maketitle

\begin{abstract}

Neural networks are commonly known to be vulnerable to adversarial attacks mounted through subtle perturbation on the input data. Recent development in voice-privacy protection has shown the positive use cases of the same technique to conceal speaker's voice attribute with additive perturbation signal generated by an adversarial network. This paper examines the reversibility property where an entity generating the adversarial perturbations is authorized to remove them and restore original speech (e.g., the speaker him/herself). A similar technique could also be used by an investigator to deanonymize a voice-protected speech to restore criminals' identities in security and forensic analysis. In this setting, the perturbation generative module is assumed to be known in the removal process. To this end, a joint training of perturbation generation and removal modules is proposed. Experimental results on the LibriSpeech dataset demonstrated that the subtle perturbations added to the original speech can be predicted from the anonymized speech while achieving the goal of privacy protection. By removing these perturbations from the anonymized sample, the original speech can be restored. Audio samples can be found in \url{https://voiceprivacy.github.io/Perturbation-Generation-Removal/}.


\end{abstract}

\section{Introduction}

    With the rapid development of \emph{neural network} (NN) in recent years, it has become the default model used in speaker recognition \cite{snyder2017deep,desplanques2020ecapa}
    and other applications \cite{he2016deep,ning2019review}.
    In \cite{goodfellow2015explaining}, the vulnerability of NNs to adversarial attacks, through subtle perturbation on the input samples, was reported. This seminal work has initiated similar studies in adversarial attacks on speaker recognition. Successful attacks on speaker models have demonstrated their ability to mislead the models into falsely identifying a speaker as someone else \cite{zhang2023imperceptible, abdullah2021hear, 9053076, li2020universal}. These findings further promote the applications of adversarial perturbation in voice-privacy protection \cite{li2023voice, chen2024adversarial}. In these studies, the adversarial perturbation technique has shown effectiveness in concealing the speaker's voice attributes from unauthorized access with malicious intentions.

    Voice attributes play a vital role in speaker recognition applications for security and forensic analysis. The misuse of voice-privacy protection techniques might have adverse consequences. For instance, criminals can exploit these techniques to mitigate forensic analysis and circumvent penalties. In such cases, it is highly desirable to reverse the voice protection and restore the original speaker attributes within the speech. In this context, efforts have been dedicated to the development of \emph{adversarial speaker purification} techniques aiming to neutralize the influence of adversarial perturbations in speaker recognition tasks \cite{wu2023defender}. Generally, the methods can be categorized as \emph{lossy-preprocessing} \cite{chen2022towards}, \emph{adding noise} \cite{chen2022towards}, \emph{filtering} \cite{chen2022towards}, \emph{denoising} \cite{li2023unified} and \emph{generative model} \cite{bai2023diffusion}. However, existing works do not aim to fully restore the speech signal. The weakness of existing techniques is manifested in three aspects: 1) existence of residual distortion in the purified speech utterances; 2) degraded performances in automatic speech recognition (ASR), pitch extraction and other downstream tasks. All these purification methods were developed under the premise that the purifiers are unaware of the perturbation generation process. In situations where the restoration of the original speech is intended, these method fail to meet the objective.

    \begin{figure*}[t]
    \centering
    \includegraphics[scale=0.77]{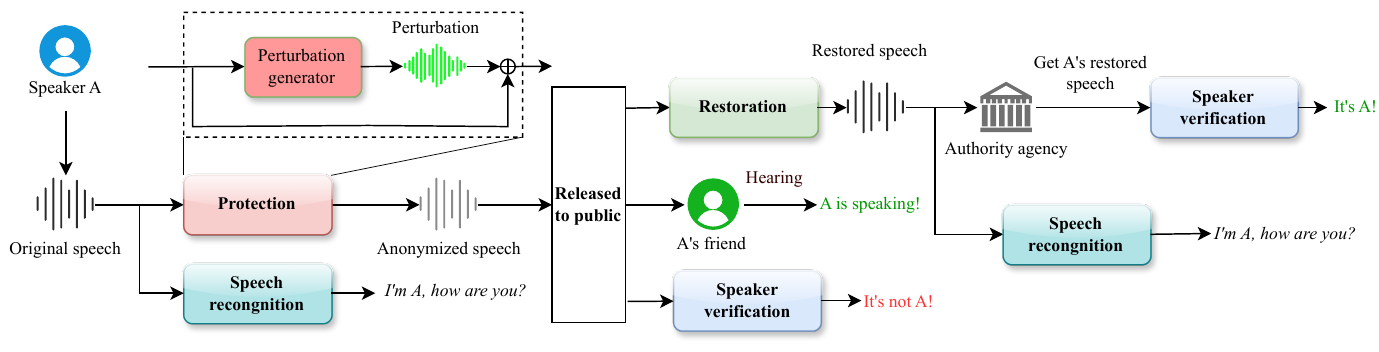}
    \caption{The process of voice-privacy protection and restoration.}
    \label{fig. system}
    \end{figure*}

    In this paper, we examine the reversibility of speaker adversarial perturbations. A well-informed (i.e., white box) setting is investigated, which represents the simplest case. More specifically, we propose a joint training framework in which the removal module is trained simultaneously with the perturbation generator and therefore well-informed on the perturbation generation process. The removal module estimates the perturbation from the adversarial speech and subtracts the estimated perturbation from the adversarial speech to restore the original speech. In this work, we adopt the symmetric saliency-based encoder-decoder (SSED) proposed in \cite{yao2023symmetric} for generating adversarial perturbations. We introduce a joint training strategy where the module responsible for removing perturbations is trained alongside the module used for their generation. The restored speech is evaluated in terms of speech quality and the efficacy of downstream tasks including speaker verification, ASR, and pitch extraction. The perturbation purification methods including the adding noise \cite{chen2022towards}, quantization (lossy-preprocessing \cite{chen2022towards}), and median smoothing (filtering \cite{chen2022towards}) methods are examined, providing a reference to our work. The experimental results obtained in speech quality, ASV, ASR, and pitch extraction evaluations validate that the joint training method could effectively restore the original speech.

    \begin{figure}[t]
    \centering
    \label{task}
    \includegraphics[scale=0.67]{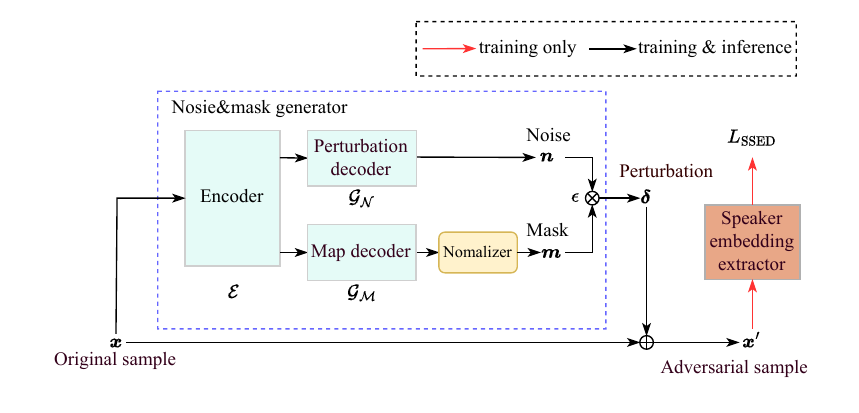}
    \caption{Framework of symmetric saliency-based encoder-decoder (SSED). The noise and mask are generated by the noise\&mask generator, represented in the rectangular box of the blue dotted line. The red lines are applicable only in training, while the black lines are valid in both.}
    \label{fig. SSED}
    \end{figure}

    \begin{figure*}[t]
    \centering
    \includegraphics[scale=0.90]{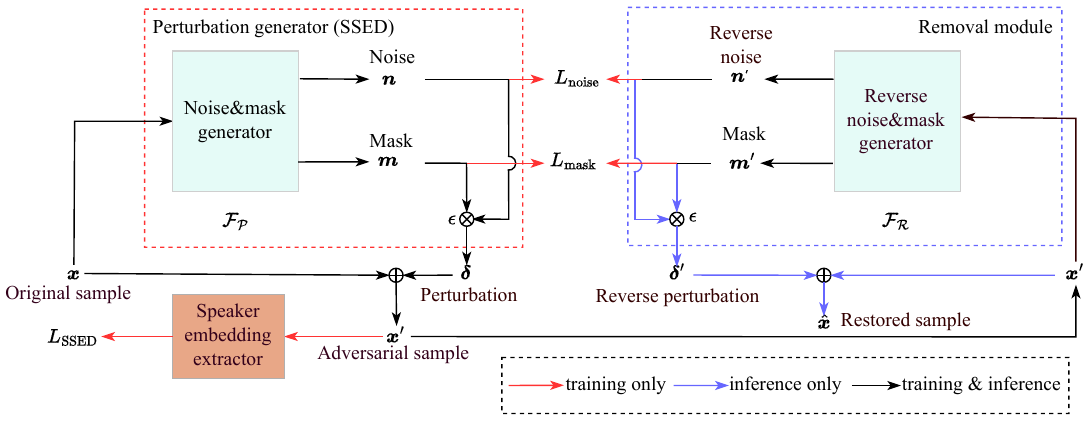}
    \caption{Model architecture for speaker adversarial perturbation generation and removal. The rectangular boxes of the red and blue dotted lines contain the perturbation generator and the removal module, respectively. The latter takes the adversarial sample ${\boldsymbol{x}}'$ generated by the former. The noise\&mask generator module used in the perturbation generator is inherited from Fig \ref{fig. SSED}. The red and blue lines are applicable only in training and inference, respectively, while the black lines are valid in both.}
    \label{fig. joint model}
    \end{figure*}

\section{Task definition}
    \label{sec. Task}
    As shown in Fig. \ref{task}, given an original speech of Speaker A, the adversarial perturbation is generated and added to it, resulting in its anonymized version. In this process, the \emph{speaker adversarial perturbation generator} is responsible for generating the perturbation. The anonymized speech can then be released and propagated, for example, through the internet, as exemplified in the figure. The voice protection application can be explained as follows: When using the anonymized utterances, individuals who are familiar with Speaker A, such as A's friend, may still be able to perceive and recognize the speaker's identity by listening. However, when subjected to a speaker recognition algorithm, the identity of Speaker A cannot be recognized.
    
    Meanwhile, consider a scenario: the anonymized speech is used as evidence for courts and public security agencies. It may be necessary to completely remove the influence of the protection to ensure the reliability of the outcomes derived from the speech utterance. In more comprehensive terms, it may be imperative to fully restore the original speaker's information, guaranteeing the confidence of the speaker recognition result obtained from the restored utterance. Furthermore, in regards to the speech content information, the ASR result should align with that acquired from the original utterance. To meet this goal, this paper works on the task of completely restoring the original speech. This paper assesses the restoration capability across three dimensions when compared to the original utterance: 1) speech quality, 2) the speaker attributes as perceived by speaker recognition models, and 3) the speech content and prosody.

\section{SSED}
\label{sec. SSED}
    In this section, we briefly review the symmetric saliency-based encoder-decoder (SSED) \cite{yao2023symmetric} in terms of its architecture and loss function.
    \subsection{Architecture}
        The architecture of SSED is shown in Fig \ref{fig. SSED}. It consists of an encoder $\mathcal{E}$, a perturbation decoder $\mathcal{G}_\mathcal{N}$, and a saliency map decoder $\mathcal{G}_\mathcal{M}$. Given the original speech $\boldsymbol{x}$, it is firstly encoded by the encoder $\mathcal{E}$ into a latent vector $\boldsymbol{y}$. Then $\boldsymbol{y}$ is decoded into a noise vector $\boldsymbol{n}$ by $\mathcal{G}_\mathcal{N}$. Parallely, $\mathcal{G}_\mathcal{M}$ is applied on $\boldsymbol{y}$, decoding it into the mask representation, which is then normalized to be the mask vector $\boldsymbol{m}$.
        The perturbation $\boldsymbol{\delta}$ is obtained by the element-wise product between $\boldsymbol{n}$ and $\boldsymbol{m}$.
        Finally, $\boldsymbol{\delta}$ is added to $\boldsymbol{x}$, resulting in its adversarial form as follows:
        \begin{equation}
        \label{eq: SSED adversarial sample}
        {\boldsymbol{x}}' = \boldsymbol{x} + \underbrace{\epsilon\cdot( \boldsymbol{n} \odot \boldsymbol{m})}_{\boldsymbol{\delta}},
        \end{equation}
        where $\epsilon$ denotes the attack intensity.

        During the training of the SSED model, the adversarial sample $\boldsymbol{x}'$ is guided by a speaker embedding extractor through the attack mechanism. In inference, given an original speech utterance $\left\{{\boldsymbol{x}}_1,...,{\boldsymbol{x}}_{\it N}\right\}$ of length $\it N$, its adversarial form is obtained as $\left\{{\boldsymbol{x}}_1',...,{\boldsymbol{x}}_{\it N}'\right\}$, giving the adversarial utterance.
    
    \subsection{Loss function}
    
        Our work focuses on protecting the original speaker without the aim of misidentifying it as a target speaker. To achieve this, the angular loss computed on the speaker embedding as defined in \cite{yao2023symmetric} is adopted to supervise the adversarial perturbation generation. Given the speaker embedding extractor as presented in Fig \ref{fig. joint model}, the speaker embedding vectors are extracted from the original and adversarial speech utterances, denoted as $\boldsymbol{z}$ and ${\boldsymbol{z}}'$, respectively. The angular loss is computed as follows:

        \begin{equation}
        \label{eq: angular loss}
        L_{\rm angular} = \frac{{\boldsymbol{z}}^{\mathsf T}\boldsymbol{z}'}{\lVert \boldsymbol{z} \lVert_{\rm 2} \lVert \boldsymbol{z}'\lVert_{\rm 2}},
        \end{equation}
    
    By minimizing $L_{\rm angular}$, the speaker similarity between the adversarial and original sample is reduced, preventing the extraction of the original speaker information by the extractor.

    Besides, the loss function to preserve speech quality is defined on the adversarial sample ${\boldsymbol{x}}'$ and the mask ${\boldsymbol{m}}$, as follows:
    \begin{equation}
    L_{\rm quality} = (1-\alpha) \lVert \boldsymbol{x}'-\boldsymbol{x} \lVert_{\rm 2} + \alpha \lVert {\boldsymbol{m}} \lVert_{\rm 2}.
    \label{eq. speech quality loss}
    \end{equation}
    As shown in (\ref{eq. speech quality loss}), $L_{\rm quality}$ is composed of two terms. The first one is the Euclidean distance between the adversarial and the original sample. The second term is the L2 norm of the mask vector. Moreover, $\alpha (0<\alpha<1)$ is the weight.
    
    Overall, the following loss function is applied in our work for the SSED model training:
    
    \begin{equation}
    L_{\rm SSED} = (1-\beta) L_{\rm angular} + \beta L_{\rm quality}.\label{eq. LSSED}
    \end{equation}
    In (\ref{eq. LSSED}), the trade-off between the adversarial perturbation and the speech quality is achieved by the weight $\beta (0<\beta<1)$.

    \begin{table*}[h]
    \caption{The EERs(\%) obtained on the recordings (rec), adversarial (adv) utterances, purified utterances and restored utterances (rst). Adding noise
    (an) was used purification method. With respect to the enrollment-test trial configurations, five types of tests are included, i.e., \emph{rec-rec}, \emph{rec-adv}, \emph{adv-adv},\emph{rec-rst}, \emph{rec-an}. The results obtained on the test-clean and dev-clean datasets are presented. Both the white-box and black-box evaluations are included.}
    \label{tb. protection result}
    \centering
    \begin{tabular}{c|c|c c c| c c|c c c |c c}
    \hline
    \multirow{3}{*}{Dataset} &\multirow{3}{*}{Gender} & \multicolumn{5}{c|}{White-box (ECAPA-TDNN)} & \multicolumn{5}{c}{Black-box (ENSKD \cite{truong2024emphasized})}\\
    \cline{3-12}
    && \multicolumn{3}{c|}{Protection} & \multicolumn{2}{c|}{Restoration}&\multicolumn{3}{c|}{Protection} & \multicolumn{2}{c}{Restoration}\\
    \cline{3-12}
    &&rec-rec & rec-adv & adv-adv &rec-rst & rec-an &rec-rec & rec-adv & adv-adv &rec-rst & rec-an\\ 
    \hline
    \multirow{2}{*}{test-clean}& male &1.21&18.62&13.69&1.21&4.82&1.21&12.17&9.56&1.15&3.78\\
    & female & 1.64&9.27&18.92&1.64&4.80&2.23&11.78&14.55&2.29&5.57\\
    \cline{1-2}
    \multirow{2}{*}{dev-clean} & male & 0.50&15.01&13.12&0.50&1.29&0.60&7.62&9.34&0.60&1.29\\
    & female &2.48&23.16&17.09&2.42&7.90&2.59&15.23&13.36&2.77&6.60\\
    \hline
    \end{tabular}
    \end{table*} 
    
\section{Proposed method}

    \label{sec. removal}
    In this section, given the SSED architecture, we propose a framework whereby the modules responsible for generating and removing perturbations are trained jointly.

    \subsection{Architecture}
        The proposed joint-training framework is shown in Fig \ref{fig. joint model}. Given a sample from the original speech $\boldsymbol{x}$, firstly, the noise\&mask generator block in SSED is applied as $\mathcal{F}_\mathcal{P}$ to generate the noise and mask vectors $\boldsymbol{n}$ and $\boldsymbol{m}$, respectively. Then, the element-wise product between $\boldsymbol{n}$ and $\boldsymbol{m}$ is computed to be the perturbation $\boldsymbol{\delta}$. The adversarial sample ${\boldsymbol{x}}'$ can be obtained by adding $\boldsymbol{\delta}$ to $\boldsymbol{x}$. Thereafter, a reverse noise\&mask generator $\mathcal{F}_\mathcal{R}$ is proposed which takes the adversarial sample ${\boldsymbol{x}}'$ as input, and is used to predict the reverse of the noise $\boldsymbol{n}$ as produced by ${\mathcal F}_{\mathcal P}$, denoted as ${\boldsymbol{n}}'$. Meanwhile, the mask vector ${\boldsymbol{m}}'$ is predicted which is expected to match $\boldsymbol{m}$. The product of ${\boldsymbol{n}}'$ and ${\boldsymbol{m}}'$ is computed to be $\boldsymbol{\delta}'$ and is expected to be the reverse of $\rm \boldsymbol {\delta}$. Finally, the \emph{perturbation removal function} is fulfilled by adding $\boldsymbol{\delta}'$ to the adversarial sample ${\boldsymbol{x}}'$, giving the restored form of the original sample $\hat{\boldsymbol{x}}$. Mathematically, $\hat{\boldsymbol{x}}$ is obtained as:
        
        \begin{equation}
            \label{eq: restored x}
            \hat{\boldsymbol{x}}={\boldsymbol{x}}' + \underbrace{\epsilon\cdot({\boldsymbol{n}}' \odot {\boldsymbol{m}}')}_{{\boldsymbol{\delta}}'},
        \end{equation}
        where $\epsilon$ denotes the attack intensity. In our work, the same intensity value is applied for generating both the adversarial perturbation and the reversed perturbation. Besides, in our architecture, $\mathcal{F}_\mathcal{R}$ and $\mathcal{F}_\mathcal{P}$ share the same structure.

    \subsection{Loss function}

        
        Additionally, guided by the expectation that $\boldsymbol{n}'$ is reverse to $\boldsymbol{n}$, a loss function is defined between them as follows: 
        \begin{equation}
        L_{\rm noise} = \lVert \boldsymbol{n} + {\boldsymbol{n}}'\lVert_{\rm 2}\label{eq. noise}.
        \end{equation}
        Meanwhile, the L2 loss is computed on $\boldsymbol{m}$ and $\boldsymbol{m}'$ as follows:
        \begin{equation}
         L_{\rm mask} = \lVert \boldsymbol{m} - {\boldsymbol{m}}' \lVert_{\rm 2}.\label{eq. norm mask loss}
        \end{equation}
        Combining (\ref{eq. noise}) and (\ref{eq. norm mask loss}), the reverse perturbation loss (rpt) is defined as follows: 
        \begin{equation}
            L_{\rm rpt} = (1-\gamma) L_{\rm mask} + \gamma L_{\rm noise},\label{eq. Lrpt}
        \end{equation}
        where the weight $\gamma (0<\gamma<1)$ balances the influence of $L_{\rm mask}$ and $L_{\rm noise}$. Finally, the training loss of the joint-training framework in Fig \ref{fig. joint model} is defined as follows:
        \begin{equation}
            L = (1-\theta) L_{\rm SSED} + \theta L_{\rm rpt}.
            \label{eq. loss}
        \end{equation}
        In (\ref{eq. loss}), the variable $\theta (0<\theta<1)$ is the weight to balance between $L_{\rm SSED}$ and $L_{\rm rpt}$.

\section{Experiments}
\label{sec. exp}

\subsection{Datasets}
    Our experiments were conducted on the LibriSpeech corpus \cite{panayotov2015librispeech}. Specifically, the train-clean-100, train-clean-360, and train-other-500 partitions were used for training. The test-clean and dev-clean datasets were used for evaluation. The recordings were resampled to 16kHz. Following \cite{yao2023symmetric}, the models worked on waveform samples. 

\subsection{Speaker embedding extractor}
    In our experiments, the speaker models adopted the ECAPA-TDNN \cite{desplanques2020ecapa} architecture, which were trained on the Voxceleb1 \cite{nagrani2017voxceleb} and 2 \cite{chung2018voxceleb2} datasets, using the open-source toolkit ASV-subtools toolkit\footnote{\url{https://github.com/Snowdar/asv-subtools}} \cite{tong2021asv}. The MUSAN corpus \cite{snyder2015musan} and the RIR datasets \cite{ko2017study} were applied for data augmentation.
    
\subsection{Compared methods}
    \noindent
    \textbf{Proposed method:} In our experiments, the perturbation generator and removal module adopted the same structure as derived from \cite{yao2023symmetric}. During training, the encoder of the ECAPA-TDNN model was adopted as the speaker embedding extractor. The weight values in the loss function were: $\alpha =0.01$, $\beta = 0.007$, $\gamma = 0.8$, $\theta = 0.06$. The learning rate was $10^{-4}$ and the attack intensity $\epsilon$ was $0.05$. The model was trained with $30$ epochs.

    \noindent
    \textbf{Purification methods:} Three speaker adversarial purification methods were examined for reference, including adding noise (SNR $= 25$) \cite{chen2022towards}, quantization (quantized factor  $=2^8$) \cite{chen2022towards}, and median smoothing (kernel size $= 3$) \cite{chen2022towards}, abbreviated as \emph{an}, \emph{qt} and \emph{ms} in the following, respectively. These experiments were performed on adversarial speech generated using the proposed jointly trained perturbation generator. Our purification code refers to the open-source toolkit \footnote{\url{https://github.com/speakerguard/speakerguard}}.

    \begin{table}[h]
    \caption{The PESQ, SNR, WERs(\%) and pitch correlation (P-Mean for mean, P-Std for standard deviation) results on the recordings (rec) the restored utterances (rst) and the purified utterances. Three purification methods are included: adding noise (an), quantization (qt), and median smoothing (ms). The results are presented on the test-clean dataset.}
    \label{tb. PESQ&SNR results}
    \centering
    \begin{tabular}{p{1.4cm}|p{0.5cm} p{0.7cm} p{0.7cm} p{0.7cm} p{0.7cm} p{0.7cm}}
    \hline
    \multirow{2}{*}{Evaluation} & \multirow{2}{*}{rec}& \multirow{2}{*}{adv}& \multirow{2}{*}{rst} &\multicolumn{3}{c}{Purification}\\ 
    \cline{5-7}
    &  &  &  & an & qt  & ms \\
    \hline
    PESQ & 4.50 & 3.67 & \textbf{4.47}& 3.15& 3.26 & 3.31 \\
    SNR & $\infty$ & 32.44 & \textbf{50.29}& 24.15 & 26.96 & 15.68 \\
    \hline
    WER & 4.08 & 6.94& \textbf{4.08} & 5.01& 5.17 & 5.61 \\
    \hline
    P-Mean & 1.00 & 0.99& \textbf{1.00} & 0.95 & 0.98 & 0.95\\
    P-Std & 0.00 & 0.03 &\textbf{0.01}& 0.11 & 0.43 & 0.08  \\
    \hline

    \end{tabular}
    
    \end{table}

\subsection{Evaluations} 

    In our ASV evaluations, both the white-box and the black-box evaluations were conducted. In the white-box evaluation, the encoder of the ECAPA-TDNN model was used to extract speaker embedding vectors. In the black-box evaluation, the open-source speaker encoder proposed in \cite{truong2024emphasized} was utilized for speaker embedding extraction \footnote{\url{https://github.com/ductuantruong/enskd}}, abbreviated as \emph{ENSKD} in this paper. Cosine distance was adopted as the speaker similarity measurement. Equal error rate (EER) was adopted as the metric.
    Our ASV evaluations were carried out in a gender-independent manner, with 20 male and 20 female speakers in both the test-clean and dev-clean datasets. 1 target and 30 nontarget trials were composed for each speaker.
    
    \noindent
    \textbf{Voice-privacy protection:} The ability of the adversarial perturbation to protect speaker information was examined with ASV evaluations. With respect to the enrollment-test trial configurations, three conditions were examined, i.e., rec-rec, rec-adv, and adv-adv. Here, \emph{rec} and \emph{adv} are short for recording and adversarial speech, respectively. Table \ref{tb. protection result} shows the results. Compared to rec-rec, rec-adv and adv-adv achieved higher EER in both white-box and black-box evaluations. This indicates that the perturbation was effective in protecting the speaker attributes within the original recordings.

    \noindent
    \textbf{Speaker information restoration:} 
    The assessment of speaker attribute restoration was conducted in the ASV evaluation using the original recordings (rec) for enrollment. The utterances restored using our proposed joint training method (rst) were used as test. Besides, results obtained on the utterances purified by adding noise (an) are presented as they achieved the lowest EERs in our experiments among the examined purification methods. The EERs are shown in Table \ref{tb. protection result}. The EERs in both the rec-an and rec-rst tests are lower compared to rec-adv, showing their efficacy in reducing the influence of the adversarial perturbation on speaker attributes. However, the rec-an evaluations still got higher EERs than rec-rec, inferring that the adding noise purification method was not able to restore the speaker attributes within the speech. Moreover, the utterance restored by the proposed method achieved similar EERs with the recordings in both white- and black-box tests, indicating that the speaker information got restored.

    \noindent
    \textbf{Speech quality restoration:} The speech quality was measured with perceptual evaluation of speech quality (PESQ) \cite{hu2007evaluation} and SNR. In our evaluations, PESQ adopted the range from $-0.5$ to $4.5$. The values were computed between the original and the test utterances. The results obtained on the adversarial and restored utterances are given in Table \ref{tb. PESQ&SNR results}. The results show that the purification methods yield lower PESQ and SNR values than adversarial speech, suggesting a degradation in speech quality by these methods.
    The restored speech achieved a PESQ score near the maximum of 4.5 and an SNR value of 50, demonstrating its effectiveness in removing the adversarial perturbation from the speech signal.

    \noindent
    \textbf{ASR evaluation:} The Whisper service \footnote{\url{https://github.com/openai/whisper}} from OpenAI \cite{radford2023robust} was called for ASR evaluation. 
    The word error rates (WERs) are given in Table \ref{tb. PESQ&SNR results}. From the table, we can see that the adversarial utterances got higher WERs than the recordings, implying the impact of the adversarial speaker perturbations on speech content. Three purification methods (adding noise and median smoothing) reduced the WER, though they did not fully attain the recording's level. However, the restored utterances obtained the same WER as the recordings. This demonstrates the effectiveness of the perturbation removal function in eliminating the influence of perturbations on speech content.

    \noindent
    \textbf{Pitch correlation:}
    Finally, the perturbation removal function in our proposed framework was assessed for its ability to restore prosody information, with pitch serving as the indicator. The pitch extractor in the VPC 2022 toolkit \footnote{\url{https://github.com/Voice-Privacy-Challenge/Voice-Privacy-Challenge-2022}} was used to calculate the pitch correlation. The statistics (mean and standard deviation) of the pitch correlation were computed as the evaluation metrics. First, for each utterance, the pitch correlation between the recording and the corresponding test utterance was calculated. Then, the mean and standard deviation were calculated across all the utterances in the test set. A higher mean and lower deviation indicate better pitch preservation. The results are included in Table \ref{tb. PESQ&SNR results}, 
    It can be observed that the purification method can damage prosody information. 
    Particularly, the mean values being 1 and deviation values close to 0 obtained by the restored speech demonstrated the effectiveness of the proposed perturbation removal function in restoring the prosody information in speech utterances.

\section{Conclusions\&future work}
\label{sec. conclusion}

This paper focuses on reversibility of the voice-privacy protection through speaker adversarial perturbation. A well-informed scenario is considered where modules for speaker adversarial generation and removal modules are trained jointly. 
Our experiments evaluated the restored speech's quality and its effectiveness in downstream tasks, including speaker verification, speech recognition, and pitch estimation. The findings showed that the original speech quality could be restored, and the perturbations' impacts on downstream tasks could be eliminated. Meanwhile, three existing perturbation purification methods were examined where the purifiers were ignorant of the perturbation generation process. Such methods demonstrated an inability to eliminate the perturbation and restore the original speech. Future work will explore the effectiveness on out-of-domain datasets and investigate the capability of denoising methods in perturbation removal.

-------------------------------------
\bibliographystyle{IEEEbib}
\bibliography{myref}
\end{document}